\documentclass[aps, prd, twocolumn, superscriptaddress, nofootinbib]{revtex4}
\usepackage{graphicx}% Include figure files
\usepackage{dcolumn}% Align table columns on decimal point
\usepackage{amssymb}% outlined letters
\usepackage{amsmath,amssymb,amsfonts}
\usepackage{color}

\begin{document}

%Macros
\newcommand{\Eq}[1]{\mbox{Eq. (\ref{eqn:#1})}}
\newcommand{\Fig}[1]{\mbox{Fig. \ref{fig:#1}}}
\newcommand{\Sec}[1]{\mbox{Sec. \ref{sec:#1}}}

\newcommand{\PHI}{\phi}
\newcommand{\PhiN}{\Phi^{\mathrm{N}}}
\newcommand{\vect}[1]{\mathbf{#1}}
\newcommand{\Del}{\nabla}
\newcommand{\unit}[1]{\;\mathrm{#1}}
\newcommand{\x}{\vect{x}}
\newcommand{\ScS}{\scriptstyle}
\newcommand{\ScScS}{\scriptscriptstyle}
\newcommand{\xplus}[1]{\vect{x}\!\ScScS{+}\!\ScS\vect{#1}}
\newcommand{\xminus}[1]{\vect{x}\!\ScScS{-}\!\ScS\vect{#1}}
\newcommand{\diff}{\mathrm{d}}

\newcommand{\be}{\begin{equation}}
\newcommand{\ee}{\end{equation}}
\newcommand{\bea}{\begin{eqnarray}}
\newcommand{\eea}{\end{eqnarray}}
\newcommand{\vu}{{\mathbf u}}
\newcommand{\ve}{{\mathbf e}}

%=====================================================================
%=====================================================================
%=====================================================================

%\title{The deformed momentum measure picture}

\title{Rainbow gravity and scale-invariant fluctuations}

\newcommand{\addressImperial}{Theoretical Physics, Blackett Laboratory, Imperial College, London, SW7 2BZ, United Kingdom}
\newcommand{\addressRoma}{Dipartimento di Fisica, Universit\`a “La Sapienza”
and Sezione Roma1 INFN, P.le A. Moro 2, 00185 Roma, Italia}

\author{Giovanni Amelino-Camelia}
\affiliation{\addressRoma}
\author{Michele Arzano}
\affiliation{\addressRoma}
\author{Giulia Gubitosi}
\affiliation{\addressRoma}
\author{Jo\~{a}o Magueijo}
%\email{magueijo@ic.ac.uk}
\affiliation{\addressImperial}
\affiliation{\addressRoma}

\date{\today}

\begin{abstract}
We re-examine a recently proposed scenario where the deformed dispersion relations associated with a flow of the spectral dimension to a UV value of 2 leads to a
scale-invariant spectrum of cosmological fluctuations, without the need for inflation. In that scenario Einstein gravity was assumed. The theory displays a wavelength-dependent
speed of light but by transforming to a suitable ``rainbow frame'' this feature can be removed, at the expense of modifying gravity. We find that the ensuing rainbow gravity theory is such that gravity switches off at high energy (or at least leads to a universal conformal coupling). 
This explains why the fluctuations are scale-invariant on all scales: there is no horizon scale as such. For dispersion relations that do not lead to exact scale invariance we find instead esoteric inflation in the rainbow frame. We argue that these results shed light on the behaviour of gravity under the phenomenon of dimensional reduction. 
\end{abstract}


\maketitle

%=====================================================================
%=====================================================================
%=====================================================================

\section{Introduction}\label{intro2}
In a recent paper~\cite{dimred} we examined the spectrum of cosmological fluctuations (scalar and tensor) expected under the following modified dispersion
relations (MDR):
\be\label{ddr1}
E^2=p^2(1+(\lambda p)^{2\gamma}).
\ee
Our calculation assumed Einstein gravity and that these MDR were realized within a higher order field theory with higher than second order spatial derivatives only.
With these assumptions we found that $\gamma=2$ leads to an exactly scale-invariant spectrum without the need to appeal to inflation. The fluctuations are produced
by a mechanism analogous to that of varying speed of light/sound models~\cite{csdot,bim,mag}.

This is very interesting because that particular dispersion relation ($\gamma=2$) is associated with a running of the spectral dimension to $d_S=2$ in the UV (see for
example~\cite{HLspec,visser}). There is mounting evidence, from a variety of fields, supporting the presence of such UV dimensional
reduction~\cite{Lollprl,Litim,Reuter,Hl,HLspec,Alesci:2011cg,Benedetti:2008gu,Modesto1,Caravelli,Magliaro,Modesto2,Calcagni1,Calcagni2,visser}.
More generally:
\be\label{ds}
d_S=1+\frac{D}{1+\gamma},
\ee
(we shall assume throughout that the number of spatial dimensions $D$ is 3).  For $\gamma=2$ scale-invariance is realized universally,
regardless of the background equation of state~\cite{dimred}.  The scale-invariance is also present while the modes are inside the horizon, being preserved as they leave it.
It is important to understand better the origin of this pervasive scale-invariance, and how it depends on the various assumptions made in the calculation.

If these MDR are valid, then we have a frequency-dependent speed of light.  However we can undo the effect by redefining the units,
for example by changing the unit of time used {\it at a given energy scale}, or by changing the units of momentum. As usual, such redefinitions merely shift
the non-trivial effects elsewhere. Sometimes the ensuing picture is awkward enough to justify sticking to the original choice of units.
Other times the equivalent picture brings new insights into the phenomenon under study, presenting a useful ``dual'' description,
as we hope will be the case in this paper.

The purpose of this paper is to transform the calculations in~\cite{dimred} into a system of units in which $c$ becomes constant. This is the so-called
rainbow metric frame~\cite{rainbowDSR}. In so doing we no longer have Einstein gravity (valid in the original units, the ``Einstein frame'').
The proposed operation is very similar to transforming from the Einstein frame (with a varying $G$) to the Jordan frame (with a constant $G$) in Brans-Dicke theory. Our purpose is to characterize gravity in the rainbow metric frame, hoping to gain intuition into the results obtained in~\cite{dimred}, and more generally into the phenomenon of dimensional reduction at high energies. As we will see $d_S=2$ is once again very strongly singled out.

\section{Rainbow gravity and linearizing variables}
``Linearizing variables'' or ``linearizing units'' are those that render the dispersion relations trivial. Usually they shift the non-trivial effects to the interactions (which are only ``minimal'' in the original units). A simple way to change variables so as to obtain non-deformed MDR is to change the time variable. Given the invariance of $x^\mu p_\mu$ under such changes of units we can infer that this is equivalent to choosing a new unit for the energy~\cite{DSRposition}. Specifically, since $x^\mu p_\mu=Et-\vect{p}\cdot \vect{x}$, a ``linearizing time'' (leaving $\vect{x}$ unchanged) must correspond to rearranging Eq.~(\ref{ddr1}) as $E^2f^2(E)=p^2$, and then defining new units $\tilde E=Ef(E)$ and $\tilde t=t/f(E)$. In the UV, $\tilde E=(E/\lambda^2)^{1/3}$. The new time unit is energy-dependent, signaling the appearance of a rainbow metric.

In the context of bimetric theories (where there is a light cone for matter and another for gravity) such a change of time variable is equivalent to transforming
from the ``Einstein'' to the ``matter'' frame. The trick was used in~\cite{piazza,ngbim} to derive the cubic action for theories with a varying speed of sound~\cite{csdot,bim} and their non-Gaussianities. The idea is to replace the standard conformal time $\eta$ by a ``fixed-$c$'' time $\tau$, such that $d\tau=c_s(\eta) d\eta$. This is a disformal transformation and in the new units the tachyonic mass term describing the Jean's instability carries the information which was previously present in the speed of sound profile.

The same thing can be done here, but with a couple of crucial differences.  Most  importantly the  ``matter frame'' is now momentum/energy dependent. Indeed the light cone of matter, as seen in the Einstein frame, depends on its frequency. This sets up a rainbow metric, as proposed in~\cite{rainbowDSR}. Since modes are labelled by a comoving $k$ (so that the physical momentum is $p=k/a$) the rainbow metric is both $k$ and time-dependent.
At high energies ($\lambda p\gg 1$):
\be\label{cofE}
c\approx {\left(\frac{\lambda k}{a}\right)}^\gamma .\ee
Therefore by changing the time unit so as to make $c$ constant,  we are necessarily disformally transforming not to a single frame, but to a class of frames labeled by $k$.

Furthermore a direct application of~\cite{piazza,ngbim} would suggest
$d\tau= c\, d\eta$.
Since $c=c(k,\eta)$ this is a non-closed form ($d^2\tau\neq 0$) and so the coordinate $\tau$ would not exist. We can either live with this (i.e. work with a non-coordinate basis) or define instead:
\be
\tau=\int c(k,\eta)\, d\eta
\ee
so that:
\be
d\tau=c\, d\eta + {\left(\int \frac{\partial c}{\partial k}d\eta\right)}\, dk.
\ee
The full Jacobian of the transformation should then be considered in all calculations.

We note that setting up a rainbow metric may 
have its problems. One may ask, for example, with respect to which 
momentum is the metric to be defined?
There are a variety of answers to this question, 
but we note that these potential ambiguities are not present in 
the context of our paper. 
We are working within the framework of linearized cosmological 
perturbation theory, so it is completely clear what the wavenumber $k$ 
means when we define the rainbow metric.

\section{The quadratic action in the rainbow frame}
We now examine the linearized fluctuations' dynamics, including gravity, as seen in the new frame. Let $\zeta$ be the curvature fluctuation, so that the quadratic action in the Einstein frame (i.e. using $\eta$-time) is:
\be
S_2=\int d^3k\, d\eta\,a^2 \left[\zeta'^2 + c^2 k^2 \zeta^2\right].
\ee
This leads to  the dynamical equation:
\be\label{veq} v''+\left[c^2 k^2 -\frac{a''}{a}\right]v=0, \ee
with $\zeta=-v/a$ (notice that the usual variable $z$ here is just the expansion factor $a$). The action for $v$ is the standard Minkowski action
when the modes are ``inside the horizon''
(i.e. when the first term in $v$ in  (\ref{veq}) dominates), and it is in terms of $v$ that we
should proceed with second quantization and evaluate the vacuum fluctuations.

We can write the action for curvature perturbations in terms of the new time variable $\tau$ taking into account the full
Jacobian of the transformation (which however boils down to a single derivative
anyway). The result is:
\be
S_2=\int d^3k \, d\tau \,y^2 \left[{\left(\frac{d\zeta}{d\tau}\right)}^2 + k^2 \zeta^2\right]
\ee
with
\be
y=a\sqrt c.
\ee
The equation of motion is now:
\be
\frac{d^2 v}{d\tau^2}+\left[k^2 -\frac{1}{y}
\frac{d^2 y}{d\tau^2}\right]v=0
\ee
with:
\be
\zeta=-\frac{v}{y}.
\ee
Also here, the action for $v$ is the standard Minkowski action 
when the modes are
``inside the horizon'', and the calculations of the vacuum fluctuations in that
regime follow through as before. But we stress that the relation between
$\zeta$ and $v$ has also changed by changing to the new frame.

\section{Scale-invariance and conformal invariance for  $\gamma=2$}
In the rainbow frame we obtain a remarkable insight into the $\gamma=2$ case,  and why it leads to scale-invariance inside and outside the horizon,
and for all equations of state~\cite{dimred}.
We find that for all equations of state, when $\lambda p\gg 1$, we have:
\be
y=a\sqrt{c}\approx \lambda k
\ee
since $c=(\lambda k/a)^2$ (cf. Eq.~(\ref{cofE})).  This means that the variable $y$ controlling Jean's instability becomes time-independent, so that the
usual mass term, $y''/y$, disappears. The perturbation equation in the rainbow frame {\it for all equations of state} becomes
the free harmonic oscillator equation:
\be\label{veqnew} \frac{d^2 v}{d\tau^2}+k^2 v=0
\ee
i.e. the theory does not know about expansion and so does not care about horizons.

This is actually what happens in the standard (Einstein) theory for radiation (by which we mean undeformed radiation)
as a result of its conformal invariance. Then $a\propto \eta$ and so the mass term vanishes:
\be
\frac{z''}{z}=\frac{a''}{a}=0
\ee
signalling the conformal invariance of radiation. But for $\gamma=2$, as seen in the rainbow frame, this happens for
all equations of state. Everything seems to be conformally invariant in the rainbow frame, and so fails to feel the effects of gravity
in a flat Friedmann model (which is conformal to Minkowski space-time).  Why this is the case remains to be better understood,
but it could be an important insight into gravity subject to dimensional reduction, with UV spectral dimension 2.

 Why do we end up with a scale-invariant spectrum? This may seem surprising because for undeformed radiation it does {\it not} happen. With undeformed radiation and Einstein gravity, for all modes $v$ is normalized as $1/\sqrt{k}$, and so we end up with $n_s=-1$. The situation is different here due to a remarkable property. Although $y$ is a time-independent, it is still $k$-dependent. Therefore:
\be\label{zetaofk}
\zeta=-\frac{v}{\lambda k}
\ee
i.e. the field redefinition which takes us from $\zeta$ to $v$ (and a Minkowski-like action in the right limit) is $k$-dependent. Consequently, even though:
\be
v=\frac{e^{-ik\tau}}{\sqrt{2k}}
\ee
we have:
\be
\zeta=\frac{e^{-ik\tau}}{\sqrt 2\lambda k^{3/2}}
\ee
i.e. a scale-invariant spectrum,
\be
\zeta=\frac{1}{\sqrt 2\lambda k^{3/2}}.
\ee
in the regime $k\eta\ll 1$.

Our findings  explain the pervasive scale-invariance found for $\gamma=2$. Firstly it is found inside and outside the horizon because in the UV (when fluctuations are produced) there are no horizons, since gravity drops out of the picture. Secondly, the transformation relating the observable $\zeta$ and the variable $v$ (subject to standard second quantization) becomes non-local in the rainbow frame. Specifically the action for $\zeta$ is:
\be
S_2=\int d^3k \, d\tau (\lambda k)^2\left[{\left(\frac{d\zeta}{d\tau}\right)}^2 + k^2 \zeta^2\right]
\ee
which becomes
\be
S_2=\int d^3k \, d\tau \left[{\left(\frac{dv}{d\tau}\right)}^2 + k^2 v^2\right]
\ee
via (\ref{zetaofk}). However this field ``redefiniton'' is $k$-dependent,
i.e. it involves gradients:
\be
v\sim \lambda \nabla \zeta
\ee
unlike in the usual case. This brings about scale-invariance
 in $\zeta$, which is
now a non-local field theory (as opposed to $v$).

\section{Dual ``inflation'' for $\gamma\neq 2$}\label{inflation}

The main objective of Ref.~\cite{dimred} was to expose a strong link 
between the choice $\gamma =2$, representative of
quantum-gravity scenarios with running of spectral dimensions to $d_S=2$ 
in the UV, and scale invariance. But it is now firmly established~\cite{Planck}
that the spectrum cannot be exactly scale invariant, so we need some departures from the $\gamma =2$ picture. A particularly simple and perhaps appealing possibility for producing such departures from exact scale invariance is to have $\gamma$ slightly lower than 2, and $d_S$ slightly higher than 2 
(as examined in~\cite{dimred}). It is then interesting to ask how 
$\gamma \neq 2$ would manifest itself in the rainbow frame.

As explained in~\cite{dimred}, when $\gamma\neq 2$  we 
solve the horizon problem in the Einstein frame with a varying speed
of light as long as:
\be\label{horsol}
-\frac{1}{3}<w<\frac{2\gamma-1}{3}\; .
\ee
It is then obvious that under this condition 
something unusual must  happen to gravity
in the rainbow frame, since the physical results cannot have changed,
and so gravity must be taking care of the horizon problem. For $\gamma=2$
the answer is that gravity simply dropped out of the picture, and so the
problem is resolved just like in Minkowski space-time. For $\gamma\neq 2$
we find that for all equations of state satisfying (\ref{horsol})
we have inflation in the dual frame.

This is easy to see explicitly. Since 
\be\label{aeta}
a\propto\eta^\frac{1}{\epsilon-1}
\ee
with $\epsilon=\frac{3}{2}(1+w)$, where $w=p/\rho$ is the equation of
state, we have that
\be
-\tau\propto (\lambda k)^\gamma \eta^{1-\frac{\gamma}{\epsilon -1}}
\ee
noting that if (\ref{horsol}) is satisfied, then 
the transformation changes the sign of
the time variable. Therefore the expansion factor is:
\be\label{atau}
a\propto (-\tau)^{\frac{1}{\epsilon -1 -\gamma}}
\ee
and we generally have accelerated expansion, but without an inflationary
equation of state. 

Indeed by comparing (\ref{aeta}) with (\ref{atau}) 
it appears that in the rainbow frame we feel an
``effective'' equation of state given by:
\be\label{neww}
{\tilde w}=w-\frac{2}{3}\gamma,
\ee
in the sense that the universe expands as if it were under Einstein gravity,
but with this matter content. 
This formula is completely general, but in the observationally relevant case
of $\gamma$ slightly smaller than 2 we find:
$
{\tilde w}\approx w-\frac{4}{3}.
$
As we see $w<1$ is necessary to generate effective inflation in the 
rainbow frame ($\tilde w<-1/3$). 
It is curious that radiation ($w=1/3$) leads to effective
de Sitter expansion ($\tilde w=-1$) in the rainbow frame.

This ``esoteric'' inflation in the rainbow frame is not new in the literature~\cite{Garattini}.
Its most interesting feature is that it may be realized purely by rainbow gravity, rather than violations of the standard energy conditions.
Its existence could lead to a whole new set of questions, namely why is it that MDRs do solve the flatness problem? This is far from obvious in the Einstein frame; yet it is evident in the rainbow frame.

We add that (\ref{ddr1}) is the most general MDR generating a power-law speed of light profile in the Einstein frame, and constant-$w$ inflation in the dual frame. MDRs that do not fall under (\ref{ddr1}) can lead, in the dual frame, to inflation with a varying equation of state, such as intermediate or graduated inflation~\cite{inter1,inter2,inter3}. A straightforward adaptation of the argument above shows that if $E^2\approx p^2g^2(p)$ in the UV, we obtain an effective equation of state in the rainbow frame:
\be
{\tilde w}= w-\frac{2}{3}\frac{g_{,p}p}{g}.
\ee
This generalizes (\ref{neww}), but if $g(p)$ is not a power-law the second term is not a constant (and it depends on $\rho$ via $a$, and on $a$ via $p=k/a$), leading to ${\tilde w}={\tilde w}(\rho)$.  We defer to a future publication a more thorough investigation of these scenarios.

\section{Conclusions}
%{\color{blue} GAC COMMENTS: $d_s=2$ in the UV as
%a preferred option for quantum gravity could be stressed more forcelly, and we could play again the card of using
%the MDR as a "probe theory" for RSD with $d_s=2$ in the UV; }

In this paper we have obtained interesting insights into gravity's behaviour under UV dimensional reduction. Our insights were obtained on the basis of a specific cosmological calculation, and therefore are a hint based on a test case, rather than proof. Yet, the suggestion seems to be that UV running to $d_s=2$ leaves all matter conformally coupled and/or blind to the effects of gravity, in the frame where the dispersion relations are rendered trivial. In fact this applies to gravitons as well (see the discussion in~\cite{dimred}), so these also do not feel the background expansion in the rainbow frame. How general this result is remains to be  investigated.

The discovery that matter is {\it always} conformally coupled to gravity
in the dual frame
sheds light on our results in~\cite{dimred}, where we found scale invariance 
inside and outside the horizon, without the usual transfer function (such as in inflation, where the spectrum goes from $n_s=-1$ to $n_s=1$ as it leaves the horizon). The spectra inside and outside the horizon are the same because the theory is conformally flat and so there is no horizon. However, in the standard case where this happens (undeformed radiation with Einstein gravity), this leaves us with the wrong spectrum, $n_s=-1$. Not so here, because the transformation between the observable $\zeta$ and the variable $v$ subject to standard second quantization is $k$ dependent in the UV, and so non-local. The $k$-dependence is such that it leaves the spectrum in $\zeta$ scale invariant, even if that of $v$ is not.

When $\gamma$ is not exactly 2 the situation is different. As we pointed out in~\cite{dimred}, the observed departures from scale-invariance~\cite{Planck}  could be explained by a very long UV transient to the asymptotic value of 2. But it could also be that the UV asymptotic value of $d_S$ is slightly higher than 2, with $\gamma$ slightly smaller than 2. Then, as we showed in Section~\ref{inflation}, we find that we have accelerated expansion in the rainbow frame even without an inflationary equation of state. Furthermore we are able to produce a near scale-invariant spectrum even without being near a deSitter phase.

% THIS IS ACTUALLY A GREAT EXAMPLE TO ADD TO THE DISCUSSION ON DIM-LESS AND OBSERVABILITY
%is frequency dependent, and so
%the effect can be expressed as a dimensionless ratio (the ratio of
%two speeds at different frequencies). Nonetheless, it is possible
%to undo the effect by redefining the units, for exemple by changing the
%unit of time used {\it at a given energy scale}. As usual, such
%redefinitions merely shift
%the non-trivial effects elsewhere. Sometimes the ensuing picture
%is complicated enough to justify the original choice of units, and the
%picture that followed. Other times the new, equivalent picture brings
%new insights into the phenomenon under study.

\section{Acknowledgments}
GAC, MA, and GG were supported in part by the John Templeton Foundation.  The work of MA was also supported by the EU Marie Curie Actions through
a Career Integration Grant.
JM was funded by STFC through a consolidated grant and by an International Exchange Grant from the Royal Society.

\bibliography{refsMOM}

\end{document}